# Insights into the Interactions between Surfactin, Betaines and PAM: Surface Tension, Small-Angle Neutron Scattering and Small-Angle X-ray Scattering Study

Jingwen Xiao,[1,†] Fang Liu,[1,†] Vasil M. Garamus,[§] László Almásy,[‡] Ulrich A. Handge,[+] Regine Willumeit,[§] Bozhong Mu,[†] and Aihua Zou[*,†]

[†]Shanghai Key Laboratory of Functional Materials Chemistry and State Key Laboratory of Bioreactor Engineering, East China University of Science and Technology, Shanghai 200237, People's Republic of China,

[‡]Wigner Research Centre for Physics, Institute for Solid State Physics and Optics, Budapest 1525 POB 49, Hungary

[§]Helmholtz-Zentrum Geesthacht, Centre for Materials and Coastal Research, Institute of Materials Research, Max-Planck-Strasse 1, 21502 Geesthacht, Germany

[+]Helmholtz-Zentrum Geesthacht, Centre for Materials and Coastal Research, Institute of Polymer Research, Max-Planck-Strasse 1, 21502 Geesthacht, Germany

_______________________________

[*] To whom correspondence should be addressed.
Tel/Fax.: +86-21-64252231.
e-mail: aihuazou@ecust.edu.cn.
[†] East China University of Science and Technology.

‡ Institute for Solid State Physics and Optics.

§ Helmholtz-Zentrum Geesthacht.

**ABSTRACT:** The interactions between neutral polymer polyacrylamide (PAM) and the biosurfactant Surfactin and four betaines, $N$-dodecyl-$N,N$-dimethyl-3 -ammonio-1-propanesulfonate (SDDAB), $N$-tetradecyl- $N,N$-dimethyl-3-ammonio-1-propanesulfonate (STDAB), $N$-hexadecyl-$N,N$- dimethyl-3-ammonio-1-propanesulfonate (SHDAB), and $N$-dodecyl-$N,N$-dimethyl-2-ammonio-acetate ($C_{12}BE$), in phosphate buffer solution (PBS) have been studied by surface tension measurements, small-angle neutron scattering (SANS), small-angle X-ray scattering (SAXS) and rheological experiments. It has been confirmed that the length of alkyl chain is a key parameter of interaction between betaines and PAM. Differences in scattering contrast between X-ray and neutrons for surfactants and PAM molecules provide the opportunity to follow separately the changes of structure of PAM and surfactant aggregates. At concentrations of betaines higher than CMC (critical micelle concentration) and $C_2$ (CMC of surfactant with the presence of polymer), spherical micelles are formed in betaines and betaines/PAM solutions. Transition from spherical to rod-like aggregates (micelles) has been observed in solutions of Surfactin and Surfactin/SDDAB ($\alpha_{Surfactin} = 0.67$ (molar fraction)) with addition of 0.8 wt% of PAM. The conformation change of PAM molecules only can be observed for Surfactin/SDDAB/PAM system in PBS. Viscosity values follow the structural changes suggested from scattering measurements i.e., gradually increases for mixtures PAM $\rightarrow$ Surfactin/PAM $\rightarrow$ Surfactin/SDDAB/PAM in PBS.

## 1. Introduction

Polymer/surfactant mixtures are used in a wide range of commercial and industrial applications[1-4], particularly, in enhanced oil recovery (EOR)[5]. Interaction between surfactant and polymer can lead to changes such as the viscosity[6], surface tension[7], conductivity[8], critical micellar concentration and aggregation number of surfactant[9] , also the surface activity of the polymer[2,10], and for some combinations the interactions are even controllable thus the desired properties can be achieved[11,12]. Therefore, polymers are always added to the surfactant systems to control rheology and stability, and to manipulate surface adsorption.

Since the coformulation of surfactant and polymer always brings about advanced or new functions, their behavior both in the bulk and at interfaces is of much current interest studied by neutral reflectometry, isothermal titration calorimetry (ITC) and etc.[13-16] Among a great deal of fundamental investigations between different polymers and surfactants, most studies have been launched on complexes involved one type of versatile polymer, the nonionic, water soluble polyacrylamide (PAM)[17-23]. For instance, Mya et al[24] studied the interaction behavior between PAM and the nonionic surfactant Triton X-100 (TX-100) and found that at higher concentrations the surfactant molecules induce a significant increase in the size of the polymer chains, while two polymers with different molecular weights interact with the surfactant quite similarly. Hai et al[25] investigated the interaction between anionic surfactant sodium dodecyl sulfate (SDS) and PAM by characterizing the microenvironment, and found that the microviscosity of the aggregate-polymer interface is greater than that of free micelle-water, which suggests that the headgroups of SDS adsorbed on PAM are more tightly packed than free micelles. Even a newly powerful and promising surfactant, Gemini has also been used to investigate the interaction with modified polyacrylamide recently[26-29]. However, to date, there have been few investigations of the interaction between PAM and a biosurfactant, not so much as the ternary complex system of PAM, a biosurfactant and amphoteric surfactant. Additionally, although the interaction between surfactants and polymers have been investigated by several methods such as NMR Spectroscopy[15], rheology[23], and conductivity[26], none of these directly give the size and even conformation information. While, small-angle neutron scattering (SANS) and small-angle X-ray scattering (SAXS) are the effective techniques capable of giving the size and shape of the aggregates.

The biosurfactant used in this work, Surfactin, lipopeptide excreted by various strains of *Bacillus subtilis*, consists of a peptide loop of seven amino acid residues (Glu-Leu-D-Leu-Val-Asp-D-Leu-Leu) and a C15 β-hydroxy hydrophobic fatty acid chain (Figure 1(a)), is extremely powerful due to its special amphiphilic character, also less toxic and easy to biodegrade, and has powerful foaming and emulsification ability[30,31], which are favorable for oil recovery. Considering the important role of PAM[32-34] and Surfactin the research in our group has been carried out the interfacial behavior between the ternary aqueous system of dodecyl betaine/lipopeptide/hydrolyzed polyacrylamide (HPAM) with crude oil[35], which revealed that the synergistic effect between biosurfactant and betaine can reduce the interfacial tension to an ultra-low level in alkaline environment, and an excellent ASP (Alkaline-Surfactant-Polymer) flooding system involving biosurfactant was obtained. In this paper the betaines and Surfactin mixtures associate with PAM, respectively were further studied, mainly with SANS and SAXS provided the interaction behavior and configuration in the bulk phase. Also the viscosity changes of the mixed systems when Surfactin and Surfactin/SDDAB was added to PAM solution by rheological experiments (flow curve measurements) were reported.

## 2. Materials and Methods

**2.1 Materials.** Surfactin was produced by *Bacillus subtilis* TD7 cultured in a laboratory of East China University of Science and Technology.[36,37] Surfactin-C15 isoform (Figure 1(a)) was separated by extraction with anhydrous ether, isolated with normal pressure ODS C18 column and purified by the RP-HPLC (Jasco, Japan). The structure of the isolated lipopeptide was determined by the electrospray ionization-time-of-flight mass spectrometer (ESI-TOF MS/MS) and GC/MS. In the pH=7.4 PBS, the -COOH groups in L-Glu1 and L-Asp5 of Surfactin were deprotonated because the $pK_a$ values of Asp and Glu are around 4.3 and 4.5, respectively.[38]

Betaines: *N*-dodecyl-*N,N*-dimethyl-3-ammonio-1-propanesulfonate (SDDAB), *N*-tetradecyl-*N,N*-dimethyl-3-ammonio-1-propanesulfonate (STDAB) and *N*-hexadecyl-*N,N*-dimethyl-3-ammonio-1-propanesulfonate (SHDAB) were purchased from Nanjing Robiot Co, Ltd., *N*-dodecyl-*N,N*-dimethyl- 2-ammonio-acetate ($C_{12}BE$) was from Zhixin Chemical Co. Ltd.. All of them were over 99% pure and were used as received.

Polyacrylamide (PAM) was purchased from Sigma-Aldrich (Shanghai) Trading

Co. Ltd. with average molecular weight of around 5,000,000-6,000,000 and used as received.

D$_2$O (99.9% D) was purchased from Sigma-Aldrich.

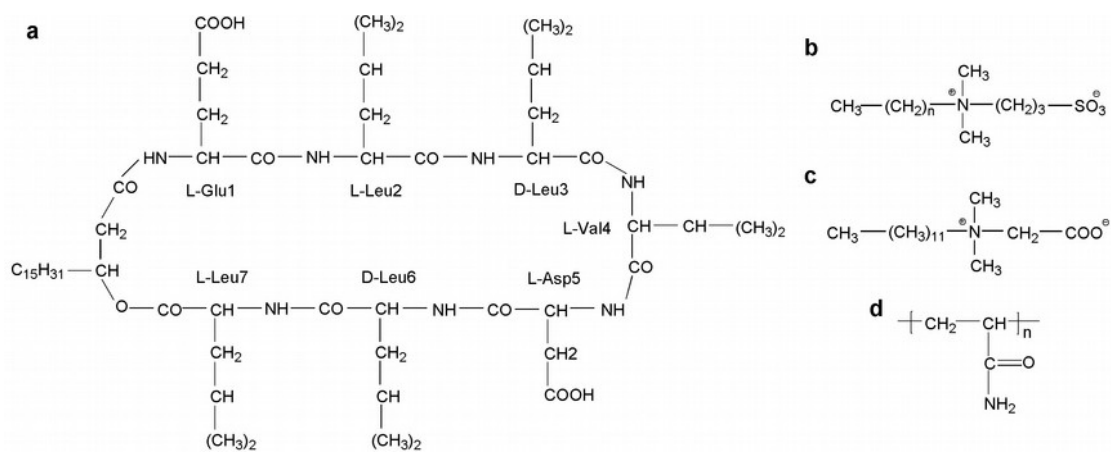

**Figure 1.** The chemical structure of Surfactin, betaines and PAM: (a) Surfactin; (b) sulfopropyl betaines, SDDAB (n=11), STDAB (n=13), SHDAB (n=15); (c) C$_{12}$BE; and (d) monomer unit of PAM.

**2.2 Methods.**

2.2.1 Sample Preparation. 10 mM PBS, pH 7.4, prepared by doubly distilled water was used as the solvent. In this study, a low concentration (0.8 wt %) of PAM was settled by consulting the result that when HPAM dosage is 0.4 g/L, the ASP ternary system is at the best performance[35]. The PAM solution was prepared first and then surfactants were added into it. After that, samples were properly mixed and then equilibrated for 1 day.

For SANS experiments, the samples were prepared with 10 mM deuterium phosphate buffer solution (p$D$ 7.4). The p$D$ value was mediated by eq 1, with the improvements made by Krezel et al.[39,40]

$$pD = 0.929pH \text{ meter reading} + 0.42 \qquad (1)$$

For the ternary system, the molar ratio of Surfactin between SDDAB and Surfactin was 0.67 as previously reported that the Surfactin/SDDAB mixed system shows a synergistic effect contribute by electrostatic attraction between the opposite charges, hydrophobic association and hydrogen bonds at this combination.[41]

2.2.2 Surface Tension. It was measured at 25 °C by tensiometer DCAT21 (Dataphysics, Germany) using a Wilhelmy small platinum plate of ca. 4 cm perimeter. The plate was first rinsed with deionized water and then burned to red before each measurement. The surface tension of deionized water was measured (72 ± 0.2 mN/m) at the beginning to check the instrument.

The surface tensions for aqueous solutions of pure Surfactin, SDDAB, STDAB, SHDAB, $C_{12}BE$, Surfactin/SDDAB ($\alpha_{surfactin}$=0.67), and each mixed with 0.8wt%PAM, respectively, were measured over a wide concentration range to determine the critical aggregation concentration (CAC), CMC and $C_2$ for each of them. Surface tension was measured three times and plotted as a function of concentration. The turning points in the plot are indicated in Figure 2.

2.2.3 Small-Angle Neutron Scattering (SANS). SANS measurements were performed on the Yellow Submarine instrument at the BNC in (Budapest, Hungary).[42] The overall interval of scattering vectors $q$ ranged from 0.09 to 3.1 nm $^{-1}$. The samples were kept in Hellma quartz cells with a path length of 5 mm and placed in a thermostatic holder at a temperature of $T$=25.0±0.5 °C. The raw scattering patterns were corrected for sample transmission, background, and sample cell scattering.[43] The 2-dimensional scattering patterns were azimuthally averaged, converted to an absolute scale and corrected for detector efficiency and dividing by the incoherent scattering spectra of pure $H_2O$, which was measured in a 1 mm path length quartz cell. The scattering from PBS buffer prepared in $D_2O$ was subtracted as the background.

2.2.4 Small-Angle X-ray Scattering (SAXS). SAXS experiments were performed at the P12 BioSAXS beamline of the European Molecular Biology Laboratory (EMBL) at the storage ring PETRA III of the Deutsches Elektronen Synchrotron (DESY, Hamburg, Germany) at 25 °C using a Pilatus 2M detector (1475 x 1679 pixels) (Dectris, Switzerland) and synchrotron radiation with a wavelength $\lambda = 1$ Å. The sample-detector distance was 3 m, allowing for measurements in the $q$ interval ranging from 0.1 to 4 $nm^{-1}$. The $q$-range was calibrated using the diffraction patterns of silver behenate. The experimental data were normalized to the transmitted beam intensity, corrected for non-homogeneous detector response, and the background scattering of the aqueous buffer was subtracted. The solvent scattering was measured before and after the sample scattering in order to control for eventual sample holder contamination. Eight consecutive frames comprising the measurements for the solvent, the sample, and the solvent were performed. No measurable radiation damage was detected by a comparison of eight successive time frames with 1 s exposures. An automatic sample changer for sample volume of 15 $\mu$L and a filling cycle of 20 s was used. The measurement time per sample was about 1 minute.[44]

2.2.5 Rheological Experiment. The stationary viscosity of the surfactant/polymer solutions was measured using a rotational rheometer (MCR502, Anton Paar GmbH,

Graz, Austria). Measurements at different shear rates in the interval from 0.1 s$^{-1}$ to 1000 s$^{-1}$ (from low to high shear rates) were performed in order to determine the flow curve (steady-state viscosity as a function of shear rate). The measurements were carried out using a Couette geometry (radius of the inner cylinder 13.3 mm, gap 1.1 mm) at a temperature of 30 °C. A volume of 18 mL was inserted into the Couette cell for each sample (one sample for one flow curve).

## 3. Results and Discussion

**3.1 Interaction between Alkyl Betaine and PAM**. Figure 2 and Table 1 present the CMC of SDDAB, STDAB, SHDAB, C$_{12}$BE, Surfactin, Surfactin/SDDAB, and the CAC (critical aggregation concentration), C$_2$ (CMC of surfactant with the presence of polymer) values of each solution mixed with 0.8wt% PAM, respectively. The concentration dependence of surface tension ($\gamma$) in the presence of polymer can be interpreted as follows[45]: At a certain concentration, often termed the CAC, there is an onset of surfactant to the polymer. Because of this the surfactant activity decrease and also the rate of decrease in surface tension is reduced. As the polymer is saturated with surfactant, the surfactant unimer concentration and the surface activity start to increase again and there is a lowering of $\gamma$ until the unimer concentration reaches the C$_2$, and normal surfactant micelles start to form. As can be seen from Table 1, CAC is smaller than CMC, which indicates that less free energy is required for surfactant and polymer forming aggregates in case of pure surfactant forming micelle. C$_2$ is generally slightly higher than the CMC of pure solution. This can be attributed to that fraction of surfactant molecules is bound to the polymer[46], so surfactant solution with the presence of polymer requires a higher concentration to form micelles which can change their size and shape to compare with pure surfactant solution.

The CAC is understood to be the point at which the polymer and surfactant start to form mixed aggregates in the bulk[47]. A low CAC value indicates that the surfactant easily binds with polymer, so it can be regarded as a measure of the interaction strength between polymer and surfactant. For SDDAB, STDAB and SHDAB, with the same polar headgroup but different hydrophobic chain lengths, when mixed with PAM respectively, the CAC decreases with increasing hydrophobic chain length. This effect can be explained by the fact that mainly hydrophobic interaction exists between surfactant and neutral polymer. SHDAB has a stronger interaction with PAM than SDDAB and STDAB due to its longer hydrophobic chain.

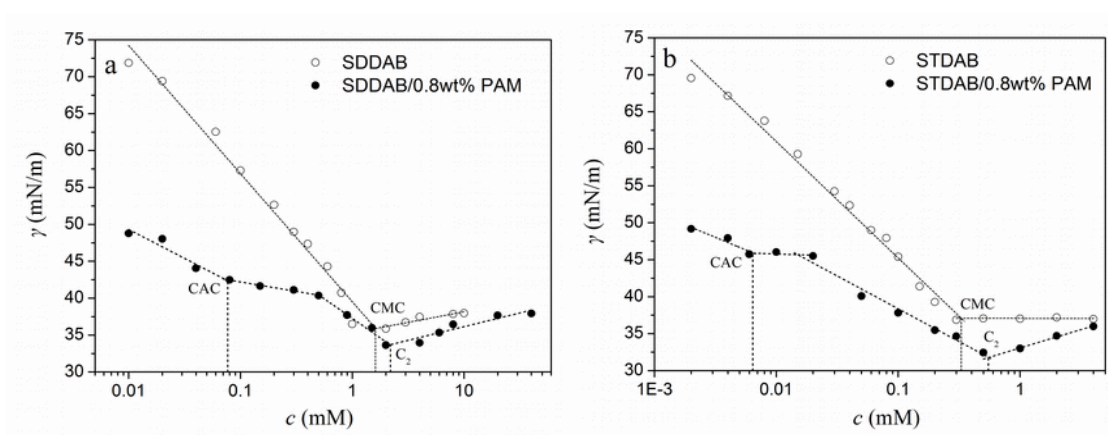

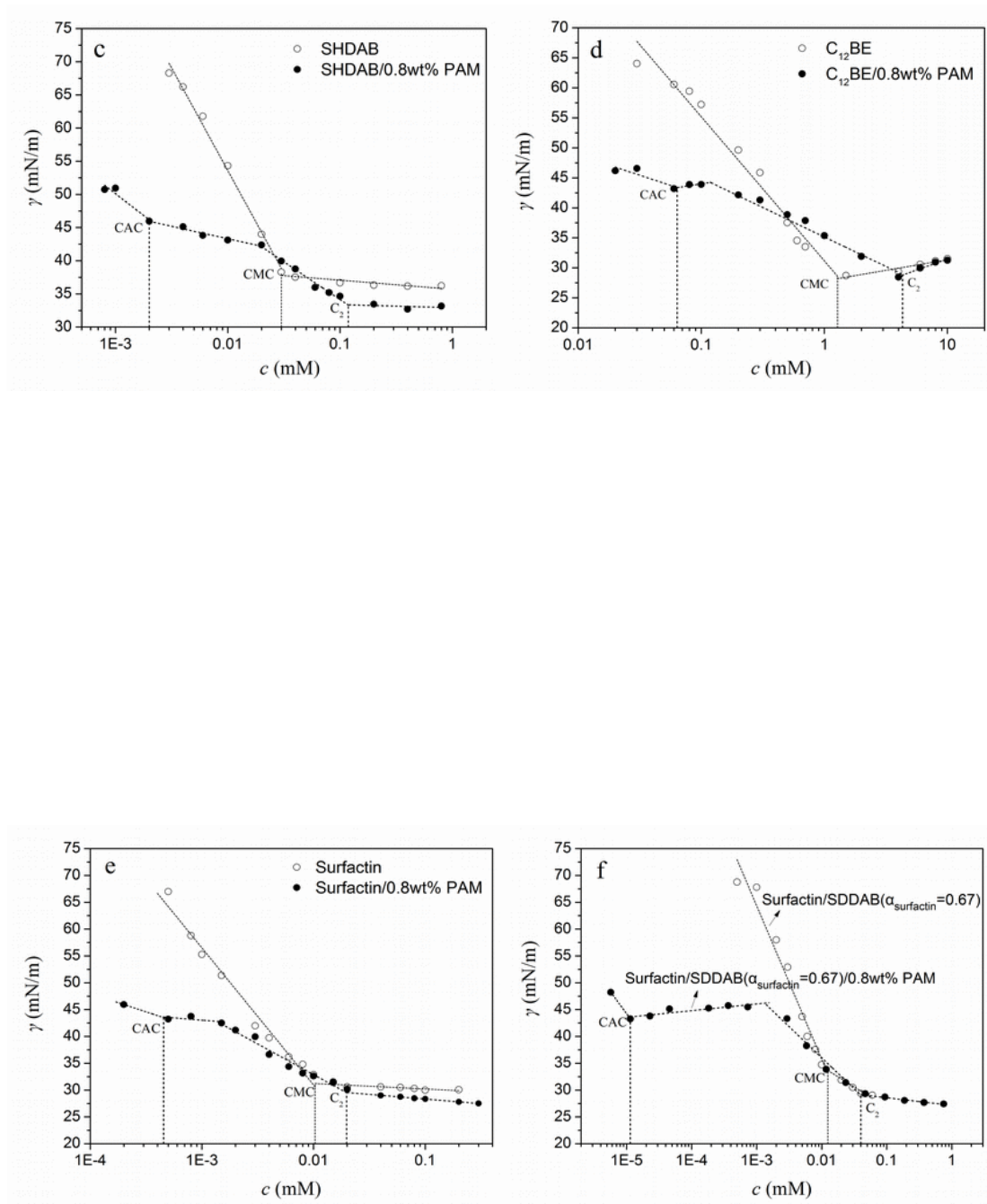

**Figure 2.** The $\gamma$-$c$ curves of betaines, Surfactin solutions and each solution mixed with 0.8wt% PAM, respectively: (a) SDDAB; (b) STDAB; (c) SHDAB; (d) $C_{12}BE$; (e) Surfactin and (f) Surfactin/SDDAB ($\alpha_{Surfactin}$=0.67).

**Table 1.   The CAC, CMC and $C_2$ Values of Surfactant Solutions and Mixed Surfactant/PAM Solutions of Different PAM Concentration (The Length of Alkyl**

| Surfactant | PAM/wt% | CAC/mM ($\pm$0.003) | CMC/mM ($\pm$0.02) | $C_2$/mM ($\pm$0.03) |
|---|---|---|---|---|
| SDDAB (C12) | 0 | | 1.61 | |
| | 0.8 | $7.68\times10^{-2}$ | | 2.22 |
| STDAB (C14) | 0 | | 0.34 | |
| | 0.8 | $6.31\times10^{-3}$ | | 0.55 |
| SHDAB (C16) | 0 | | $2.95\times10^{-2}$ | |
| | 0.8 | $1.97\times10^{-3}$ | | 0.12 |
| $C_{12}$BE (C12) | 0 | | 1.27 | |
| | 0.8 | $6.37\times10^{-2}$ | | 4.31 |
| Surfactin (C15) | 0 | | $1.03\times10^{-2}$ | |
| | 0.8 | $4.52\times10^{-4}$ | | $1.99\times10^{-2}$ |
| Surfactin/SDDAB ($\alpha_{surfactin}$=0.67) | 0 | | $1.24\times10^{-2}$ | |
| | 0.8 | $1.14\times10^{-5}$ | | $4.09\times10^{-2}$ |

As to SDDAB and $C_{12}$BE, with the same hydrophobic chain but different hydrophilic groups, both the CAC and CMC of SDDAB are larger than those of $C_{12}$BE. This can be attributed to that the sulfopropyl group of SDDAB is more hydrophilic due to a higher degree of dissociation than the carboxyl group in $C_{12}$BE. Moreover, the length of bridge chain, which connects positive charge and negative charge, is different either. SDDAB has two more methylenes than $C_{12}$BE, which leads to an increase of the dipole-dipole repulsion between the headgroups, causing the increase of CMC.[48-50] Furthermore, the polar headgroup of SDDAB is larger than that of $C_{12}$BE, which is less favorable for self-aggregation and interaction with PAM as well. As a result, compared with SDDAB, at the same concentration, $C_{12}$BE self-aggregates more readily, and the CMC is smaller and has a stronger interaction with PAM.

Figure 3(a) presents SANS data of 0.8wt% PAM and the mixed system of 0.8wt % PAM with 1 mM and 10 mM SDDAB, respectively. All of the fitted parameters are listed in Table 2. The scattering intensities ($I(q)$) are proportional to the square of volume of aggregates and of the contrast between neutron scattering length densities of the aggregates and the solvent.[51] When the surfactant concentration is smaller than $C_2$, only single surfactant molecules interact with polymer molecules (strands of polymer molecules). The scattering intensity and radius of gyration ($R_g$) of PAM does not change significantly after 1 mM ($< C_2$) SDDAB is added, suggesting that the PAM configuration does not change much. The above result is also supported by the similar *p(r)* functions, the pair distance distribution function of a particle, shown in Figure 4(a). The diagram in Figure 5(a) shows this interaction. However, when the concentration of SDDAB was increased to 10 mM, which is much larger than its $C_2$, the scattering intensity increased markedly, meanwhile Rg dropped from 67.4 Å to 25.9 Å and the maximum diameter ($D_{max}$, the upper limit for the maximum particle dimension) dropped from 190 Å to 70 Å, indicating that smaller sized but more aggregates are formed, which are slightly larger than the micelles in SDDAB solution. One can suggest that polymer molecules curl around SDDAB micelles, as indicated by the longer tail of *p(r)* function at larger *r* and the larger $D_{max}$ required for describing of SANS data in presence of PAM (Figure 4(b)). This interaction sketch is shown in Figure 5(b).

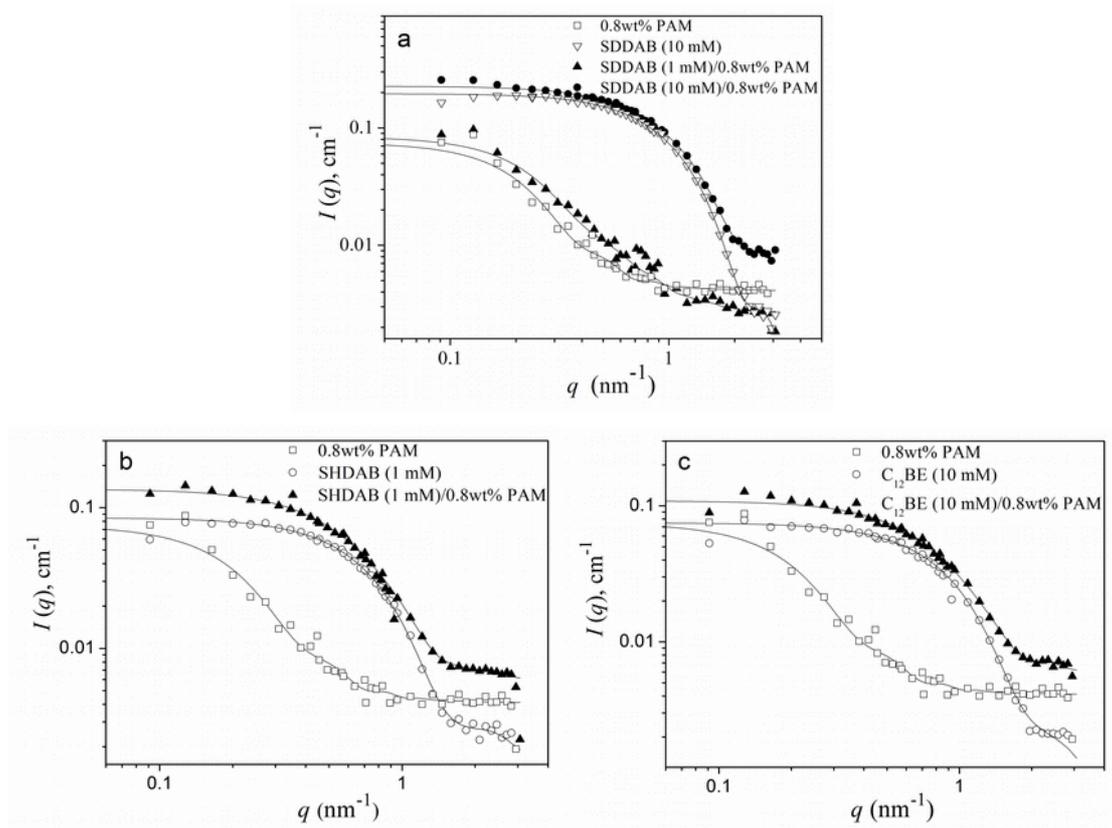

**Figure 3.** SANS data of (a) 0.8wt% PAM, 10 mM SDDAB and 0.8wt% PAM mixed with 1 mM and 10 mM SDDAB, respectively; (b) 0.8wt% PAM, 1mM SHDAB and SHDAB (1mM) /0.8wt% PAM; and (c) 0.8wt% PAM, 10 mM $C_{12}$BE and $C_{12}$BE (10 mM)/0.8wt% PAM.

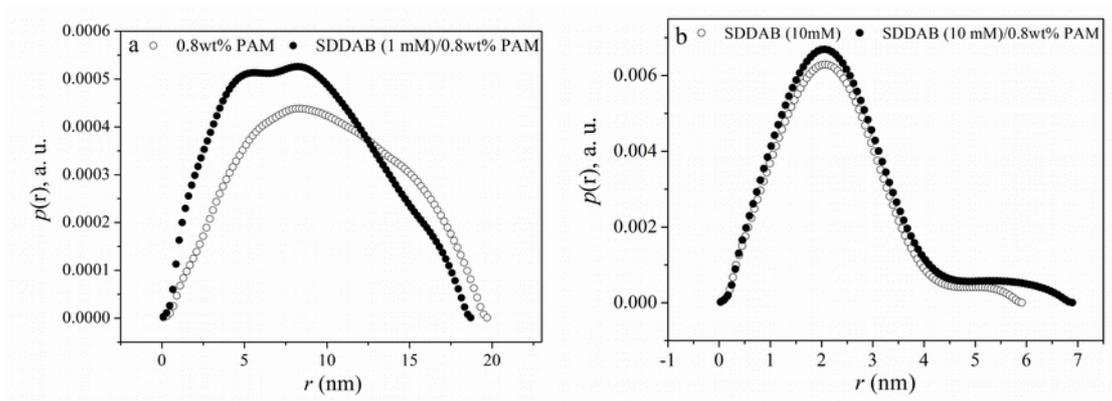

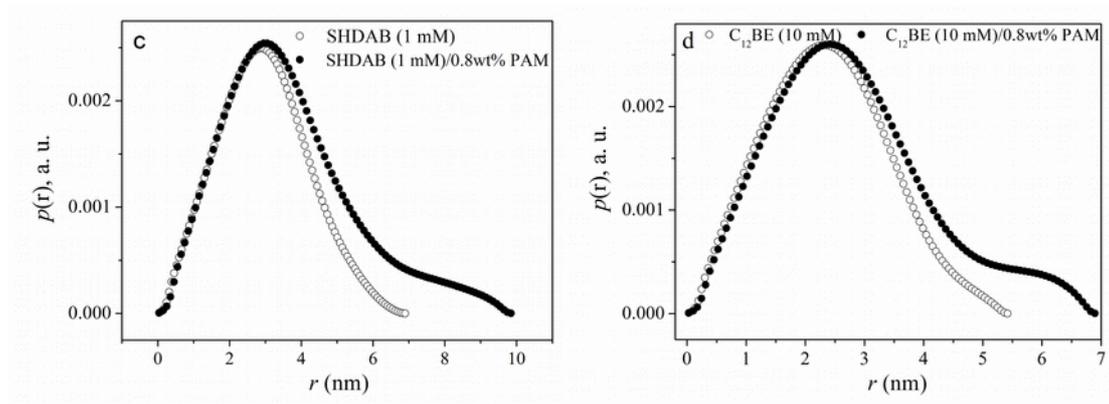

**Figure 4.** *p(r)* function of (a) SDDAB (1 mM)/0.8wt% PAM; (b) SDDAB (10 mM) 0.8wt% PAM; (c) SHDAB (1 mM)/0.8wt% PAM; and (d) $C_{12}BE$ (10 mM)/0.8wt% PAM.

**Table 2. Characteristic Parameters: $D_{max}$, $R_g$, and $I_o$ for Micelles of Surfactant and Surfactant/PAM (The Bold Items in the Table Correspond to the Values of Rod-like Micelle)**

| Surfactant | PAM/wt% | $D_{max}$/nm | $R_g$ or $\mathbf{R_{g,cs}}$ /nm | I(0) or $\mathbf{I_{CS}(0)}$/cm$^{-1}$, cm$^{-1}$nm$^{-1}$ |
|---|---|---|---|---|
| PAM | 0.8 | 20.0 | 7.65±0.25 | 0.071±0.006 |
| SDDAB (1 mM) | 0.8 | 19.0 | 6.74±0.23 | 0.082±0.005 |
| SDDAB (10 mM) | 0 | 6.0 | 1.74±0.01 | 0.197±0.002 |
|  | 0.8 | 7.0 | 2.59±0.10 | 0.155±0.008 |
| SHDAB (1 mM) | 0 | 7.0 | 2.29±0.02 | 0.083±0.001 |
|  | 0.8 | 10.0 | 2.92±0.07 | 0.130±0.002 |
| $C_{12}BE$ (10 mM) | 0 | 5.5 | 1.85±0.02 | 0.075±0.001 |
|  | 0.8 | 7.0 | 2.20±0.04 | 0.103±0.002 |
| Surfactin (0.4 mM) | 0 | 5.5 | 1.75±0.09 | 0.010±0.001 |
|  | 0.8 | **3.1** | **1.08±0.05** | **0.003±0.0001** |
| Surfactin/SDDAB (0.4mM, $\alpha_{Surfactin}$=0.67) | 0 | 5.9 | 1.95±0.08 | 0.017±0.001 |
|  | 0.8 | **3.2** | **1.07±0.04** | **0.004±0.0001** |

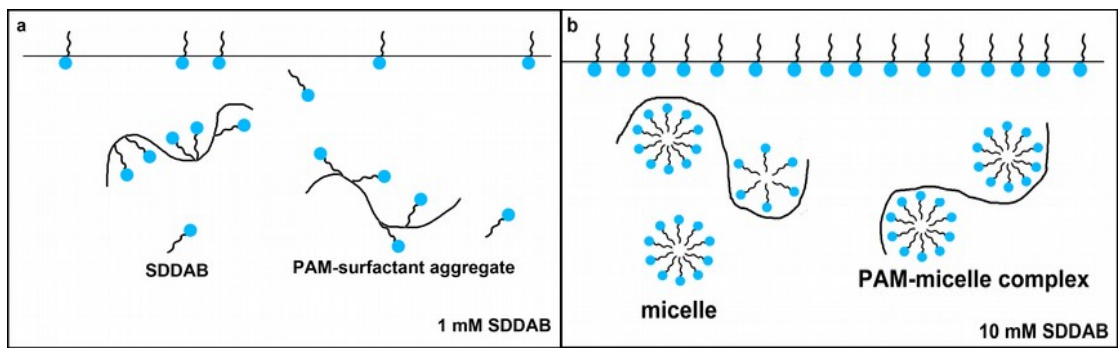

**Figure 5.** Sketch of the interaction between PAM and (a) 1 mM SDDAB; (b) 10 mM SDDAB.

Figures 3(b), 3(c) and Figure 4(c), 4(d) show the scattering curves and the corresponding $p(r)$ function of SHDAB (1 mM), $C_{12}BE$ (10 mM) in PBS buffer (the concentrations are much larger than their CMC and $C_2$), and with 0.8wt% PAM, respectively. It can be seen that both of these two pure betaines form spherical micelles. When 0.8wt% PAM is added to SHDAB (1 mM) and $C_{12}BE$ (10 mM), the shape of the micelles does not change, they remain spherical, which means there is weak interaction between betaines (SHDAB and $C_{12}BE$) and PAM. This behavior is similar to the system of SDDAB/PAM. Near spherical micelles entangled by PAM are formed.

**3.2 Interaction between Surfactin and PAM.** Figure 2(e) and Table 1 show that compared with betaine/PAM systems, Surfactin/PAM has a CAC as low as $10^{-4}$ magnitude, indicating that there is strong interaction between them. Figure 6(a) presents the SANS results of PAM, Surfactin, and Surfactin/PAM. Pure Surfactin form spherical micelles, similar to pure betaine solutions. However, in contrast to betaine/PAM systems, which all form spherical micelles, when 0.8wt% PAM is added to 0.4 mM Surfactin, whose concentration is far above its CMC, the scattering curves indicate formation of non-spherical micelles and stronger interaction. When the value of q exceeds Guinier range (q·Rg >1), and the scattering curves follow the power law, i.e. I (q) ~ q$^{-\alpha}$, the exponent α can be used as a simple parameter to determine the micelle shape: α=4 corresponding to spherical micelle, α=2 corresponding to vesicles or disc-like micelles, and α=1 corresponding to rod-like micelle.[52] The slope of I(*q*)-*q* curve of 0.4 mM Surfactin/0.8wt% PAM mixture at low q is -1, i.e. the α value is 1, indicating that rod-like micelles are formed. A sketch of this conformation transition is shown in Figure 6(c).

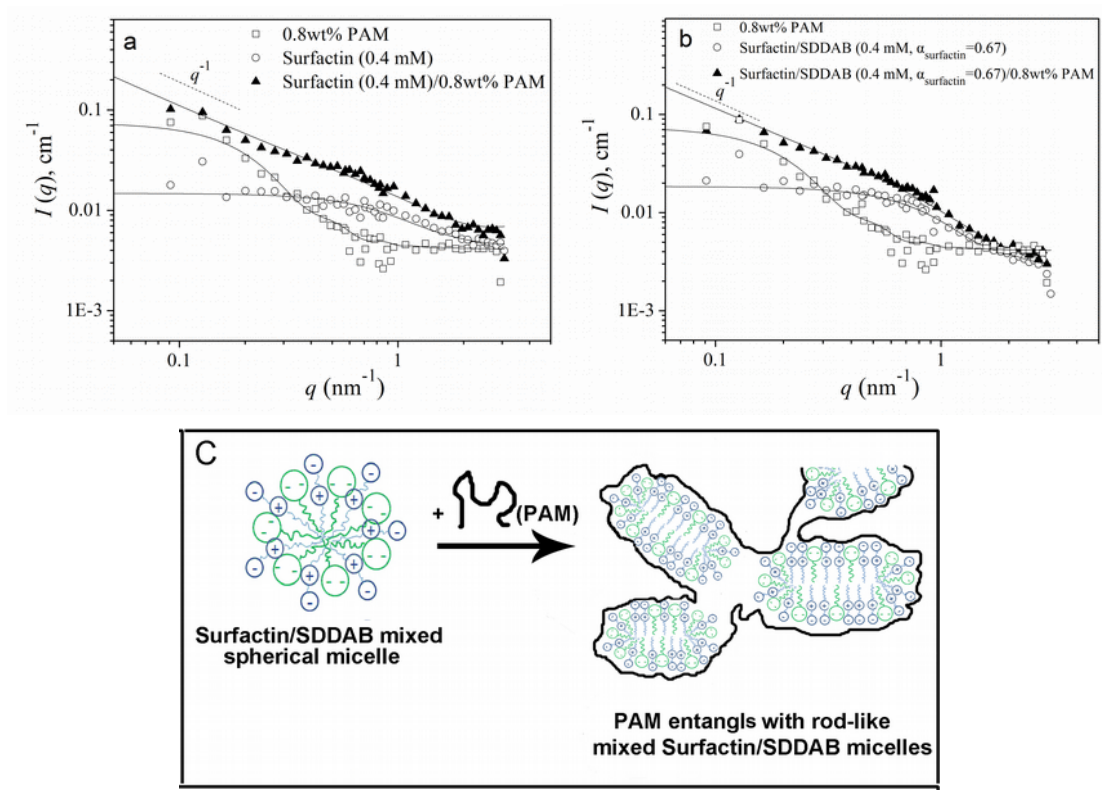

**Figure 6.** SANS data of (a) 0.8wt% PAM, Surfactin (0.4 mM) and Surfactin (0.4 mM)/0.8wt% PAM mixtures; (b) 0.8wt%PAM, Surfactin/SDDAB (0.4 mM, α$_{Surfactin}$=0.67) and Surfactin/SDDAB (0.4 mM, α$_{Surfactin}$=0.67)/0.8 wt% PAM; (c) a sketch of the aggregate conformation transition when PAM is added to the Surfactin/SDDAB mixture.

**3.3    Properties    of    Surfactin/SDDAB/PAM    Compound    System.** Surfactin/SDDAB mixture interacts with PAM most strongly since this ternary system shows the minimum CAC (Figure 2(f)). In Figure 2(f), between CAC and $C_2$, the long plateau has a slight positive slope. Staples et al[53] studied oppositely charged polymer/surfactant mixtures of SDS and poly(dimethyldiallylammonium chloride). They observed a pronounced increase in the surface tension at a concentration between the surfactant concentrations normally attributed to the CAC and $C_2$. By investigating the amount of surfactant and polymer at the interface and about the structure of the adsorbed layer, they revealed that the increase was associated with a partial desorption of polymer-surfactant complex. Also for this oppositely charged system, Campbell et al[54] attributed this increase to the comprehensive precipitation of bulk complexes into sediment. In the ternary system of Surfactin/SDDAB/PAM, no real precipitation was observe, and this slight positive slope might result from a partial desorption of polymer-surfactant complex in this ternary system due to the strong interaction between surfactant mixture with PAM[53].

Figure 6(b) presents the SANS results of Surfactin/SDDAB (0.4 mM, $\alpha_{Surfactin}$=0.67), and Surfactin/SDDAB (0.4 mM, $\alpha_{Surfactin}$=0.67)/0.8wt% PAM. Previous study has proved that Surfactin and SDDAB mixtures form spherical micelles[41], however, when PAM is added to the mixture, as the same with the Surfactin (0.4 mM)/0.8wt% PAM system, the slope of scattering curves at low q turns to -1, which means rod-like micelles are formed in this mixed system.

When SDDAB is added to Surfactin, both hydrophilic headgroups arrange more closely firstly as the charges are neutralized. After PAM is added, the hydrophobic association between the PAM chain and the hydrophobic core of the mixed surfactants micelles keeps the headgroups near each other. Under these circumstances, the headgroup of Surfactin in the shell takes the S2 configuration, which is a saddle-like structure with two charged acid residues forming a particular "claw" configuration and three hydrogen bonds [NH (7)-CO (5), NH (4)-CO (2), NH (6)-C$^1$O] which reduces the ring flexibility[55]. Thus Surfactin exists in a more compact status to reduce the steric hindrance between the headgroups. So hydrophilic headgroups are arranged more closely and the contact area with water is reduced. That is why rod-like micelles are more preferable.

In our previous study we found that with the only addition of either SDDAB or Surfactin to HPAM, the interfacial tension between crude oil and water would not be reduced to an ultra-low level, while this could be achieved by using the mixture of SDDAB and Surfactin[35]. In present paper, by comparing the interaction strength among the systems of betaine/PAM, Surfactin/PAM and Surfactin/SDDAB/PAM, it is confirmed that the surfactants mixture is most readily to associate with PAM, mainly because SDDAB is favorable for reducing the electrostatic repulsion between the negative charges of Surfactin, so the synergistic effect between them yields a more pronounced interaction with the polymer, and this is the exact type of the complex ternary flooding system which shows the best performance we have obtained before.

### 3.4 SAXS Results

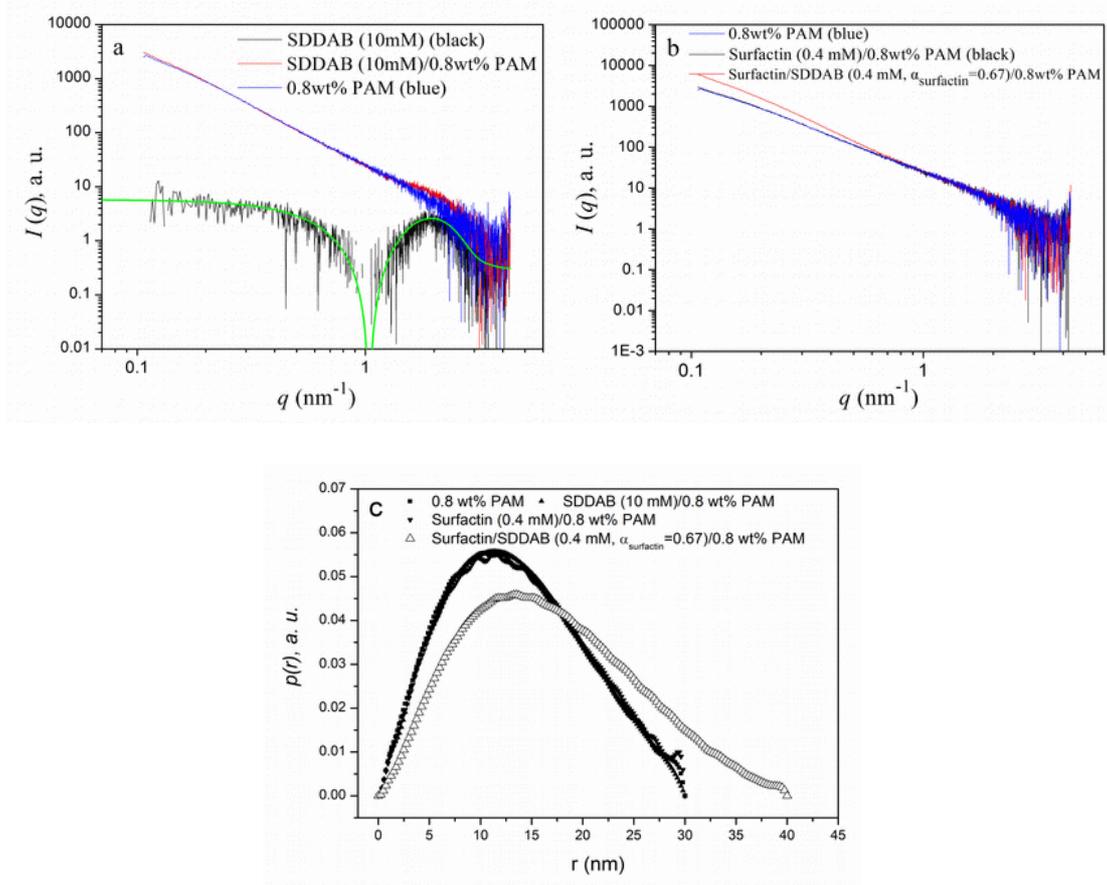

**Figure 7.** Results of SAXS measurements of surfactants, PAM and surfactant/PAM systems. (a) SAXS curves of 0.8 wt% PAM (blue); SDDAB (10 mM) (black); and (c) SDDAB (10 mM)/0.8 wt% PAM (red). Solid line (green) is a fit of IFT analysis for SDDAB (10 mM) (black). (b) 0.8 wt% PAM (blue); Surfactin (0.4 mM)/0.8wt% PAM (black) and Surfactin/SDDAB (0.4 mM, $\alpha_{\text{Surfactin}}$=0.67)/0.8wt% PAM; (c) $p(r)$ functions obtained from SAXS data in solution of 0.8% of PAM in PBS (filled squares), in 10 mM SDDAB (filled triangles up), in 0.4 mM of surfactin (filled triangles down) and in 0.4 mM Surfactin/SDDAB ($\alpha_{\text{Surfactin}}$=0.67) (empty triangle).

It can be seen from Figure 7 that SAXS scattering intensity is dominated by PAM contribution. Intensity for PAM is 2 orders of magnitude higher than that for SDDAB. In the case of SDDAB, the scattering from micelles is observed with maximal diameter of 5.5 nm and radius of gyration of 2.8±0.1 nm (results of IFT analysis, see Table S1 in Supporting Info). Shape of corresponding p(r) (Figure 7(c)) is more complicate than for neutron measurements due to different sign of scattering contrast for alkyl chain core (radius of around 1 nm) and polar group outer shell region (radius ~ 2 nm). The scattering pattern of 0.8 wt% PAM does not change after addition of either 10 mM of SDDAB or 0.4 mM of Surfactin. Only addition of the mixture of Surfactin and SDDAB brings some changes in the conformation of PAM i.e., the intensity increases at lowest q interval and the slope increases from -2 to -2.3. Indirect Fourier Analysis confirm these observations and shape of *p(r)* function changes only for solution of Surfactin/SDDAB/PAM and PAM chains transform to larger and more compact structures i.e., curles around SDDAB/Surfactin mixed micelles, in agreement with the SANS data. Obtained radius of gyration increases from 10.5 to 13.2 nm (see Table S1 in Supporting Info). Values of gyration radius for PAM aggregates obtained from SANS and SAXS should be considered with caution while complexes formed by PAM are too large for measured interval of scattering vector q and give level of lowest size of aggregates.

**3.5 Rheological Results.** Figure 8 depicts the steady-state shear viscosity as a function of shear rate for pure PAM, Surfactin and Surfactin/SDDAB mixed with PAM, respectively. At low shear rates up to approximately 10 s$^{-1}$, the viscosity does not depend on shear rate (Newtonian regime). At shear rates higher than 10 s$^{-1}$, the viscosity decreases with shear rate (shear thinning or pseudoplastic behavior). This effect is caused by the intramolecular and intermolecular associations of PAM in solution leading to self-aggregation and to the formation of larger aggregates. Above a critical shear stress molecular aggregation is disturbed and thus the viscosity is lowered. The value of the critical shear stress is in the order of 0.15 Pa - 0.18 Pa for the pure PAM solution, and in the order of 0.2 Pa after addition of Surfactin and approximately 0.25 Pa after addition of Surfactin/SDDAB, respectively. When Surfactin is added to the PAM solution, the viscosity increases due to formation of rod-like micelles and further increase of viscosity for Surfactin/SDDAB/PAM is caused by simultaneous changes in micelles (sphere-rod transition) and in polymer conformation (bulky - more compact structure).

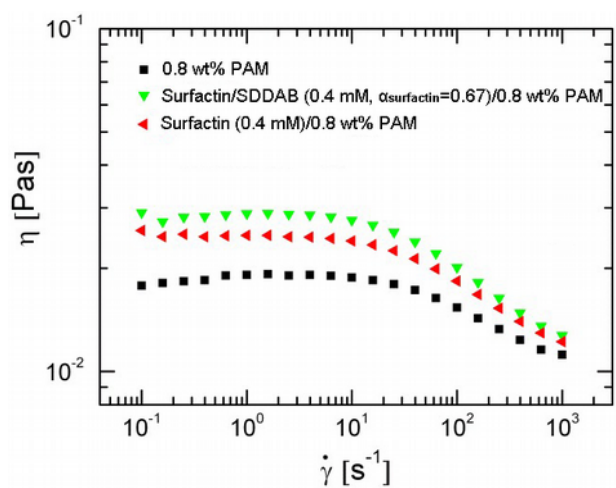

**Figure 8.** Dependence of steady-state viscosity on shear rate for solutions of PAM, Surfactin/PAM and Surfactin/SDDAB/PAM.

## 4. Conclusions

Interaction between zwitterionic betaines and neutral polymers is mainly hydrophobic, with the length of alkyl chain of surfactants playing a major role. The more hydrophilic betaine shows lower interaction strength with PAM. When the concentration of betaine is higher than its $C_2$, it forms spherical micelles (similar ones to polymer-free mixtures) which decorate PAM molecules without changing the polymer's conformation. The micelles formed by the anionic biosurfactant Surfactin show sphere-to-rod transition after addition of 0.8wt% of PAM and no influence on polymer conformation. In the surfactant mixture Surfactin/SDDAB ($\alpha_{Surfactin}$=0.67) and PAM, both the sphere-to-rod transition of the surfactants aggregates and the changes of polymer conformation have been observed. These structural changes at the microscopic level explain the special properties of Surfactin/SDDAB ($\alpha_{Surfactin}$=0.67) and PAM in PBS solution i.e., the observed highest interaction from surface tension measurements and highest viscosity from rheology measurements. Moreover, that the conformation change of PAM only happens in the ternary (Surfactin/betaine/PAM) system suggests more priority to the use of Surfactin/betains mixtures which exhibit higher effect of associating with neutral polymer. These results are favorable for the use of Surfactin/betaine mixture in the ASP flooding system with ultra-low interfacial tension[35].

## Acknowledgments

Aihua Zou gratefully acknowledges the support of this work by the Alexander von Humboldt Foundation. We gratefully acknowledge the support of this work by Chinese National Natural Science Foundation (201003047) and Fundamental Research Funds for the Central Universities. The SANS measurements have been performed with the support of the European Commission (Grant agreement N 283883-NMI3). This work was supported by project KMR12-1-2012-0226 granted by the National Development Agency (NDA) of Hungary. The experimental support of Ivonne Ternes (rheological measurements) and Clement Blanchet (SAXS measurements) is gratefully acknowledged. The great English writing improvement from Prof. Leonard I. Wiebe, Department of Oncology, Faculty of Medicine & Dentistry, University of Alberta is also gratefully acknowledged.

**Supporting Information Available**: The introduction of the indirect Fourier transformation (IFT) method we used to analyze scattering data and the fitting results from SAXS data. This material is available free of charge via the Internet at http://pubs.acs.org.

Kingswood, 1994; pp 166-175.

**Table of Contents Only:**

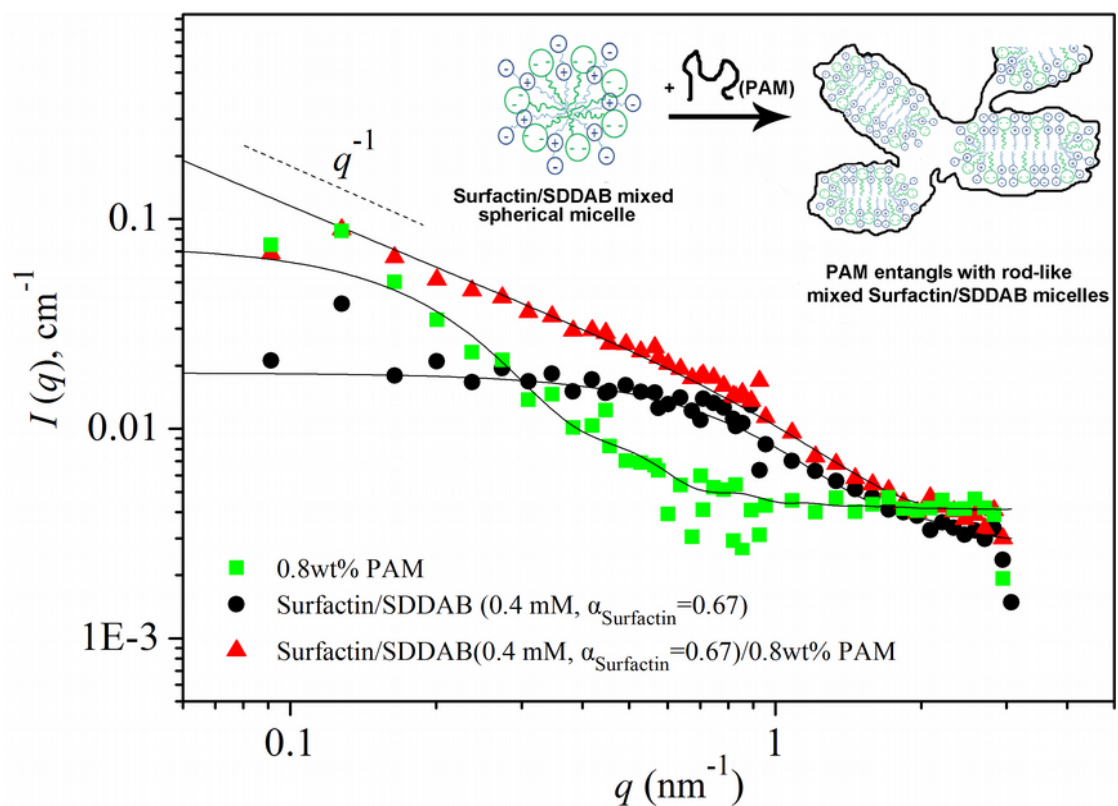

Surfactin/SDDAB mixed spherical micelle turns into rod-like micelle after PAM is added
(as suggested by SANS results)